\documentclass[12pt,a4paper,titlepage]{article}
\usepackage{amsmath,amsthm,amsfonts}
\usepackage[active]{srcltx}
\usepackage{graphicx}
\begin{document}

\newcommand{\EF}{\mathcal{E}}
\newcommand{\abs}[1]{\left\lvert #1 \right\rvert}
\newcommand{\norm}[1]{\left\lVert #1 \right\rVert}
\newcommand{\ket}[1]{\lvert #1 \rangle}
\newcommand{\brkt}[2]{\langle #1 \vert #2 \rangle}
\newcommand{\braket}[3]{\langle #1\lvert #2\rvert #3\rangle}
\newcommand{\ketbra}[2]{\lvert #1 \rangle\langle #2 \rvert}
\newcommand{\schrod}{Schr\"odinger }
\newcommand{\bs}{\negthickspace}
\newtheorem{lemma}{Lemma}
\newtheorem*{thm}{Theorem}
\newtheorem{unbounded}{Theorem}
\parskip = 0.5\baselineskip

\begin{titlepage}
\begin{center}

\vspace{1cm}  {\huge \bf The gravitational
analog of the \\
Aharonov-Bohm electric effect \vspace{1cm}}

 { \tt by}\vspace{1cm}

 {\large \bf Doron Ludwin$^1$}\vspace{1cm}

 \vspace{0.1cm}
 {\small $^1$Department of Physics, Technion, Haifa 32000, Israel}

\end{center}
\textbf{Abstract}\\
\\
The electric Aharonov-Bohm effect is a special case of the general
Ab effect. However, when inserting a gravitational potential in
the place of the time dependent potential, a different
understanding of the phase shift could be gained. The usual
topological phase is replaced by a phase with origin in the red shift of the particle 
at one of the paths taken relative to the other path. In this case, the change in the 
geometrical measure is the source of the phase shift, which therefore has a local 
interpretation along with the non-local topological explanation.

\end{titlepage}
\pagenumbering{arabic}
\section{Introduction}
The AB (Aharonov-Bohm) effect, is thought of as the main example
to the idea that the electromagnetic potential has
physical significance \cite{ab59}. The wavefunction of a charged particle which
travels through an AB potential, will receive a change is phase
which could be observed through the shift of the interference fringes in a Young double slit type experiment.

This phenomena is usually interpreted as a non-local quantum
phenomena, connected with the non-trivial topology of the AB
potential. This interpretation is associated with the
understanding that the charged particle wavefunction receives this phase change
although there is no "local" interaction with the electromagnetic
field. Normally, such an interaction would have caused momentum
transfer to the particle which would result with a change in the
particle's velocity, but to our best knowledge, such a change in
velocity hasn't been observed. Furthermore, it has been
theoretically predicted that apart from a phase change there is no
other physical phenomenon.

In further work \cite{deWitt66,Dowker67,Dowker_Roche67,Papinir67},
generalizing the AB effect to the gravitational potential,
different gravitational fields have been taken to show, that the
AB phase difference may exist between two paths taken around a
gravitational analog to the AB solenoid (such as a cosmic string),
although the particles travel on a flat curvature. This
is a gravitational analogue to the AB magnetic effect. \\
In this work, however, we develop an interesting gravitational
analog to the \textbf{electric AB effect} \cite{ac83,apv88}, which might teach us
something further concerning the AB potentials in the gravitational case.

We show that when inserting a constant gravitational potential in
the place of the time dependent potential in the electric AB
effect, the usual topological phase could be replaced by a phase
due to the red shift of the particle on one of the
path taken relative to the other path. In this case the change of the geometrical measure induces the same phase shift as given by the topological argument.

In this work, we also show that when splitting the wave function
into two parts, and expect to interfere the parts again, the role
of proper time must be taken carefully into consideration.

\section{The Electric AB effect}
\subsection{The ordinary Electric AB effect}
Let us look at a region in space where a particle moves under the
influence of a time dependent potential, $U(t)$, with zero
gradient over space.

This particle's Lagrangian is given by:
\begin{equation}
    {\cal L}=\frac{1}{2}m\dot{x}^2-U(t)
\end{equation}

Using the Euiler-Lagrange Equation:
\begin{equation}
    \frac{d}{dt}(\frac{\partial{\cal L}}{d\dot{x}})=\frac{\partial{\cal L}}{dx}
\end{equation}
we get:
\begin{equation}
    m\ddot{x}=0
\end{equation}
which means there is no net force acting upon the particle,
although the potential the particle senses changes through time.
There is naturally a change in the total particle's energy, but
this additional energy, constant over this region in space, acts
as a constant energy shift added to the system, and will not be
noticed by any measurement of a physical local parameter.

On the other hand, in QM, the Hamiltonian of the particle is given
by:
\begin{equation}
    H=H_o+U(t)
\end{equation}
where $H_o$ is the Hamiltonian of a free particle.

Taking $\Psi_o$ as the wavefunction of the free particle, and
trying a solution of the type:
\begin{equation}\label{PSI with potential}
    \Psi=\Psi_oe^{-iS/\hbar},
\end{equation}
we get that the schr\"{o}dinger equation takes the form:
\begin{equation}
    i\hbar\frac{\partial\Psi}{\partial t}=(i\hbar\frac{\partial\Psi_o}{\partial
    t}+\Psi_o\frac{\partial S}{\partial
    t})e^{-iS/\hbar}=[H_o+U(t)]\Psi=H\Psi
\end{equation}
which gives the straightforward solution:
\begin{equation}\label{the phase factor}
    S=\int U(t)dt
\end{equation}

According to this, in QM the wave function of the particle changes
when this time varying potential is introduced, only the change is
added as a phase factor, and therefore does not effect any
classical result. However, in an interference experiment, this
phase could be noticed as a phase difference in an interference
picture.

\begin{figure}
\begin{center}
  \includegraphics{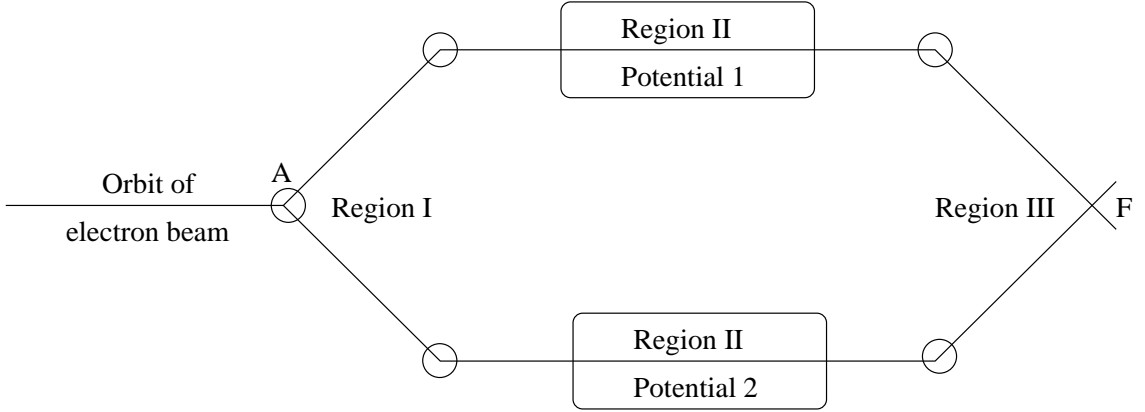}\\
  \end{center}
  \caption{Schematic experiment to demonstrate the electric AB effect with a time varying potential}\label{Electric AB}

\end{figure}

For example, consider a single electron when the wave function splits
into two parts due to some kind of barrier, and each part of the
wavefunction is allowed to enter a long cylindrical metal tube, as
shown in fig. \ref{Electric AB}

After the parts of the wave function pass through the tubes, they
are combined to interfere coherently at $F$. The electric
potential, $U(t)=e\phi(t)$, in each tube is determined by a time
delay mechanism in such a way that the potential is zero in region
$I$(until the electron's wave packet is well inside the tubes).
The potential then grows as a function of time, but differently in
each tube. Finally it falls back to zero, before the electron
comes near the other edge of the tube. Thus, the potential is
nonzero only while the electrons are well inside the tube (region
$II$). The purpose of this arrangement is to ensure that the
electron is in a time-varying potential without ever being in a
field. This is assured by the fact that the electron never
senses a space gradient in the potential, but merely a time
gradient.

Now, let $\Psi=\Psi_1^0(x,t)+\Psi_2^0(x,t)$ be the wavefunction
before the electron enters the tubes. According to (\ref{PSI with potential}) and (\ref{the phase factor}),
the wavefunction, after passing through the tubes,
is altered only by a phase factor as follows:
\begin{equation}
    \Psi=\Psi_1^0e^{-iS_1/\hbar}+\Psi_2^0e^{-iS_2/\hbar}
\end{equation}
where
\begin{equation}
    S_1=e\int\varphi_1(t)dt,\quad S_2=e\int\varphi_2(t)dt
\end{equation}

It is evident that the interference of the two parts at F will
depend on the phase difference $(S_1-S_2)/\hbar$. Thus, there is a
measurable physical effect of the potentials even though no force
is ever actually exerted on the electron. The effect is quantum
mechanical in nature since it arises in the phenomenon of
interference.

The phase difference $(S_1-S_2)/\hbar$, can also be expressed as
an integral around a closed circuit in time:
\begin{equation}
    \triangle \Phi=\frac{e}{\hbar}\oint\varphi dt
\end{equation}

The covariant statement of the above conclusion, points that there
should be a similar result involving the space part of the four
vector $A_\mu$, meaning the vector potential, $\textbf{A}$. That
gives the more well known effect known as the magnetic AB effect,
which we shall not deal with here.

\subsection{An Electric AB effect in an elevator}
We begin, as before, with an electron with wavefunction split
into two parts, each going into a different elevator in
which they continue to travel at the same speed in the $x$
direction. The elevators can travel up and down in a static
electric field, caused for instance by a great flat capacitor,
which has a constant electric field $E$, with a positive potential
gradient in the $z$ direction, but no force acting in the $x$
direction. After the particle's wave function is well inside the
elevators, the elevators climb (against the electric field force)
to a height $h$. The elevator 1 comes back down right away, while
elevator 2 stays at that height for a time $\Delta t$ and then
comes back down. The scenario is arranged this way to ensure that
the only difference between both paths will be the time $\Delta
t$, in which elevator 1 returns down, making sure the particle's
wave function has seen a constant electric potential $V_1$, while
the other elevator stays at height $h$ assuring the
particle's wave function sees a potential $V_2$.

Although the same exact forces act upon the two possible electron
paths, the time $\Delta t$ in which the particles were under
different electric potentials, with no force acting upon them,
gives rise to an additional different phase shift for each of
them:
\begin{eqnarray}
  \Psi_1&=&\Psi_{total}e^{\frac{ie}{\hbar}V_1\Delta t} \\ \nonumber
  \Psi_2&=&\Psi_{total}e^{\frac{ie}{\hbar}V_2\Delta t}
\end{eqnarray}

The phase shift is therefore:
\begin{equation}
    \Delta \phi=\frac{e}{\hbar}(V_1-V_2)\Delta t
\end{equation}

This is a very simple example in which the electric potential,
even if it stays constant over the path, has an influence on a
phase shift, which might be encountered. The reason we have discussed
the elevator is in preparation for the gravitational analog given in the next
section.

\section{The gravitational analog of the electric AB effect}
The gravitational analog to the AB effect has been discussed widely in the literature
\cite{deWitt66,Dowker67,Dowker_Roche67,Papinir67}. However, the
analogy was given to the magnetic effect, where different metrics
have been considered which have topological singularities although
the particle travels in a locally flat space with no curvature,
while the analogy to a locally curved space has not been
discussed. These authors have shown that there is an analogy
between the electromagnetic potential $A_\mu$ and the
gravitational vector
$h_\mu=(\frac{1}{2}h_{00},h_{10},h_{20},h_{30})$ for the low
velocity limit on a weak gravitational field, where $h_{\mu\nu}$
is a small curvature disturbance over the flat Minkowski space,
such that:
\begin{equation}
    g_{\mu\nu}=\eta_{\mu\nu}+h_{\mu\nu}
\end{equation}

It was shown that the expected change in phase of the wave
function of a particle of mass $m$, when taken around a curve
$V_1$ is:
\begin{equation}
    \delta\beta=\frac{m}{\hbar}\int_{V_1}h_\mu dx^\mu
\end{equation}
in complete analogy with the electromagnetic case.
\\The magnetic analog, involving the analogy between the spacial parts of the four
vectors $\mathbf{A}\sim \mathbf{h}$, has been studied
in the literature. However, the analogy to the electric effect,
which yields the analogy $\phi\sim h_0$,  hasn't been discussed.
Perhaps that is because it seems less "magical" since the particle
travels in an ordinary curved space, where gravitational forces
may act on it. However, trying to follow the influence of local
features and not merely topological features of the space-time,
the gravitational analog to the electric effect is particularly interesting.
This way, the contribution of the change in the
space-time measure caused by the curved space to the phase shift
could be quantified.

\subsection{The Newtonian analog of the effect}

We shall start by replacing the electric potential given in the
last section with a Newtonian gravitational potential, which is
given by:
\begin{equation}
    \Phi=-\frac{GM}{R}
\end{equation}
where $M$ is the mass of earth and $R$ is the distance from the
center of earth. The Newtonian gravitational constant $G$ is taken
to be 1 from now on.

Taking the particle again through the choice of two different
elevators that start at an altitude $R_1$. Both elevators climb to
an altitude $R_2$ through a similar path, where at the end, the
first elevator travels straight down, and the second elevator
stays at the altitude $R_2$ for a time $\Delta t$ and then goes
down along a path similar to the first one. At the whole time, the
particle didn't change it's vertical velocity $v$. Following
exactly the same steps as before, the only difference in phase is
caused by the time $\Delta t$ when the particle travels in the
elevators at the same speed but at different altitudes.\\

At that part of the particle's path, the Hamiltonian of the
particle is:

\begin{equation}
    H=H_o+m\Phi
\end{equation}

where $H_o$ is the Hamiltonian of a free particle, and $m\Phi$ is
constant over the time $\Delta t$.

We then get the straightforward solution:
\begin{equation}\label{psi with potential}
    \Psi=\Psi_oe^{im/\hbar \frac{M}{R}\Delta t}
\end{equation}

So although there is no force acting in the direction of motion upon the particle at that
time, the difference in the constant gravitational potential the
wave function encounters causes the change in phase:

\begin{equation}
    \Delta \phi=\frac{mM}{\hbar}\left(\frac{1}{R_2}-\frac{1}{R_1}\right)\Delta t
\end{equation}

This is the expected phase shift when introducing the
gravitational potential instead of the electric potential. Moving
on to general relativity, we shall first show that we get the
Newtonian result also for the weak gravitational field
approximation (in the low velocity limit), and then we shall see
that the phase could be explained by the difference in the
space-time measure between both paths of the particle.

\subsection{The gravitational effect in a weak Schwarzschild
field}
\subsubsection{The straightforward approach}
In the gravitational analog, we have an analogy between $\phi$ and
$\frac{1}{2}h_{00}$, where for the weak Schwarzschild field:
\begin{equation}
    h_{00}=g_{00}+1=\frac{2M}{R}
\end{equation}
According to this, the electric phase shift:
\begin{equation}
    \Delta\phi=\frac{e}{\hbar}\oint \Phi(t) dt
\end{equation}
is replaced by:
\begin{equation}
    \Delta\phi=\frac{m}{\hbar}\oint \frac{1}{2}h_{00}(t) dt=-\frac{m}{\hbar}\oint \frac{M}{R(t)}dt
\end{equation}
Thinking of our experiment as carried over a circular path of the
particle, and after omitting the identical contributions to the
phase and to the amplitude given by both parts of the path, the
only non-vanishing contributions are given by:

\begin{equation}
    \Delta\phi=-\frac{m}{\hbar}\int_{\Delta t} \frac{M}{R_1}dt - \frac{m}{\hbar}\int_{-\Delta t} \frac{M}{R_2}dt=\frac{mM}{\hbar}\left(\frac{1}{R_2}-\frac{1}{R_1}\right)\Delta t
\end{equation}

getting as expected the Newtonian result for the weak field
approximation. In the next section, we get this
result through a semi-covariant description of a wave equation,
and see what are the conditions on the wave equation to get the
correct result.

\subsubsection{A semi-covariant description of the wave equation}

Starting from the scalar product of the energy-momentum vector:
\begin{equation}
    g_{\mu\nu}p^\mu p^\nu=-m^2c^4
\end{equation}

Separating the equation to the time and spatial parts, for a block
separated metric tensor, we can get:
\begin{equation}
    -g_{tt}E^2=g_{ij}p_ip_jc^2+m^2c^4
\end{equation}
which gives:
\begin{equation}
    \sqrt{-g_{tt}}E=\sqrt{g_{ij}p_ip_jc^2+m^2c^4}=mc^2\sqrt{1+\frac{g_{ij}p_ip_j}{m^2c^2}}\approx
    mc^2 + \frac{g_{ij}p_ip_j}{2m}
\end{equation}

Replacing $\sqrt{-g_{tt}}E$ with the operator $-i\hbar\sqrt{-g^{tt}}\frac{\partial}{\partial
t}$, we get the wave equation:

\begin{equation}\label{semi-covariant wave equation}
-i\hbar\frac{\partial\Psi}{\partial
t}=\left[\frac{mc^2}{\sqrt{-g^{tt}}}+\frac{g^{ij}p_ip_j}{2m\sqrt{-g^{tt}}}\right]\Psi
\end{equation}

Equation (\ref{semi-covariant wave equation}), is a good approximation for describing a geodesic motion of a free falling spinless massive particle, in the environment of a time-space block separated metric. The form of the equation is Schrodinger-like, and assumes causal evolution in the direction of the $t$ axis (this is correct in the weak field approximation and far from metric singularities).

For the Schwarzschild metric, $-g^{tt}=(1-\frac{2M}{Rc^2})^{-1}$, and
therefore, for the weak field approximation:
\begin{equation}\label{time part}
    (-g^{tt})^{-1/2}\approx 1-\frac{M}{Rc^2}
\end{equation}

Furthermore, in the particle's path we are looking at, the
particle travels at a constant altitude $R$, and therefore,
$p_r=0$. Therefore for the Schwarzschild metric:

\begin{equation}\label{the momentum at constant R}
    g_{ij}p_ip_j=g_{rr}p_r^2+p_\Omega^2=\mathbf{p}^2
\end{equation}

Substituting (\ref{time part}) and (\ref{the momentum at
constant R}) into (\ref{semi-covariant wave equation}), we
obtain:

\begin{equation}
    -i\hbar\frac{\partial\Psi}{\partial
t}=\left[mc^2 + \frac{\mathbf{p}^2}{2m}-
\frac{mM}{R}\left(1+\frac{\mathbf{p}^2}{2m^2c^2}
\right)\right]\Psi
\end{equation}

In the low velocity limit $mc\gg \mathbf{p}$, and therefore, the
last term falls out leaving the wave equation:

\begin{equation}\label{the correct wave equation}
    -i\hbar\frac{\partial\Psi}{\partial
t}=\left[mc^2 + \frac{\mathbf{p}^2}{2m}- \frac{mM}{R}\right]\Psi
\end{equation}

This result is consistent with the Newtonian
approximation. Taking the gravitational field into account in a
path at a constant height, induces an effective
potential $U(R)=\frac{mM}{R}$, which changes the energy of the
system without changing the momentum. This is exactly the
Newtonian potential which causes the effect at the gravitational
analog of the AB electric effect. In the next section we shall see how we get this result using the role of proper time for the free-falling particle in the background of the Schwarzschild metric.

\subsection{A Space-Time local explanation of the phase shift}

We shall now try to get the effect, by the phase difference which
is a result of the space-time self length difference between both
paths.\\
For that we shall assume that each of the parts of the
wave function, travels on a locally flat metric, and the
Schwarzschild metric is noticed only when looking at the
difference between both paths in interference. We also make sure that there is a
clock which measures the proper time at each elevator to
make sure the elevators are synchronized to one another. \\
The time $\Delta t$ in which the first elevator stays at altitude
$R_2$ could therefore be given as $\Delta \tau$ (proper time) instead. If we now
look at the wave equation which describes the evolution of each
separate part of the wave-function, at different constant
altitudes, we get the ordinary Schrodinger equation for a free
particle, according to proper time $\tau$ instead of $t$:
\begin{equation}\label{proper time evolution}
    -i\hbar\frac{\partial\Psi}{\partial\tau}=mc^2+\frac{\textbf{p}^2}{2m}
\end{equation}
However, if we want the parts of the wave-function to interfere in
the end, we must go bach to the evolution according to the coordinate $t$.

The connection between $dt$ and proper time $d\tau$ is:
\begin{equation}
    d\tau^2=-ds^2=-g_{tt}dt^2-g_{rr}dr^2-R^2d\Omega^2
\end{equation}

For the low velocity limit, where the advance in time $dt$ is much
greater then the spatial advance:
\begin{equation}\label{red shift}
    d\tau^2=-g_{tt}dt^2 \quad \Rightarrow \quad d\tau=\sqrt{-g_{tt}}dt
\end{equation}
which is the result for an ordinary red shift.

Substituting \ref{red shift} into \ref{proper time evolution}, we
get the evolution for the variable $t$:
\begin{equation}
    -i\hbar\frac{1}{\sqrt{-g_{tt}}}\frac{\partial\Psi}{\partial t}=mc^2+\frac{\textbf{p}^2}{2m}
\end{equation}
or:
\begin{equation}
    -i\hbar\frac{\partial\Psi}{\partial
    t}=\sqrt{-g_{tt}}\left(mc^2+\frac{\textbf{p}^2}{2m}\right)
\end{equation}

As before, we take the Schwarzschild weak field approximation:
\begin{equation}\label{time part of metric}
    (-g_{tt})^{1/2}\approx 1-\frac{M}{Rc^2}
\end{equation}

and obtain:

\begin{equation}
    -i\hbar\frac{\partial\Psi}{\partial
t}=\left[mc^2 + \frac{\mathbf{p}^2}{2m}-
\frac{mM}{R}\left(1+\frac{\mathbf{p}^2}{2m^2c^2}
\right)\right]\Psi
\end{equation}

and as before, in the low velocity limit $mc\gg \mathbf{p}$, and
therefore, the last term falls out leaving the wave equation:

\begin{equation}
    -i\hbar\frac{\partial\Psi}{\partial
t}=\left[mc^2 + \frac{\mathbf{p}^2}{2m}- \frac{mM}{R}\right]\Psi
\end{equation}

This result is similar to (\ref{the correct wave equation}).
We can see that this equation will lead exactly to the analog of
the electric AB effect, although it has a simple alternative local explanation
to the phase shift in agreement with the topological analysis.

\section{Summary}

We have studied the gravitational analog of the electric AB effect and find that the topological argument leads to an effect that can be equivalently explained by a local change in effective path length, raising interesting questions for further study in the relativistic case.

\end{document}